\documentclass[a4paper,fleqn]{cas-dc}
\usepackage[switch]{lineno}
\usepackage{enumitem}

\usepackage[numbers]{natbib}
\usepackage{booktabs}
\usepackage{url}

\newif\ifcommenton
\commentonfalse
\ifcommenton
\linenumbers
\newcommand{\hsd}[1]{\textcolor{red}{\emph{[Shenda: #1]}}}
\newcommand{\rr}[1]{\textcolor{blue}{\emph{[Revision: #1]}}}
\newcommand{\ra}[1]{\textcolor{blue}{\emph{[Reviewer 1: #1]}}}
\newcommand{\rb}[1]{\textcolor{red}{\emph{[Reviewer 2: #1]}}}
\newcommand{\rc}[1]{\textcolor{brown}{\emph{[Reviewer 3: #1]}}}

\else
\newcommand{\hsd}[1]{}
\newcommand{\rr}[1]{}
\newcommand{\ra}[1]{}
\newcommand{\rb}[1]{}
\newcommand{\rc}[1]{}
\fi

\begin{document}
\let\WriteBookmarks\relax
\def\floatpagepagefraction{1}
\def\textpagefraction{.001}
\shorttitle{Opportunities and Challenges of Deep Learning for ECG Data}
\shortauthors{Shenda Hong et~al.}

\title [mode = title]{Opportunities and Challenges of Deep Learning Methods for Electrocardiogram Data: A Systematic Review}                      

\author[1]{Shenda Hong}
\ead{sdhong1503@gmail.com}
\author[2,3]{Yuxi Zhou}
\ead{joy_yuxi@pku.edu.cn}
\author[2,3]{Junyuan Shang}
\ead{sjy1203@pku.edu.cn}
\author[4]{Cao Xiao}
\ead{cao.xiao@iqvia.com}
\author[5]{Jimeng Sun}
\ead{jimeng.sun@gmail.com}

\address[1]{Department of Computational Science and Engineering, Georgia Institute of Technology, Atlanta, USA}
\address[2]{School of Electronics Engineering and Computer Science, Peking University, Beijing, China}
\address[3]{Key Laboratory of Machine Perception (Ministry of Education), Peking University, Beijing, China}
\address[4]{Analytics Center of Excellence, IQVIA, Cambridge, USA}
\address[5]{Department of Computer Science, University of Illinois at Urbana-Champaign, Urbana, USA}

\begin{abstract}

\noindent \textbf{Background:} The electrocardiogram (ECG) is one of the most commonly used diagnostic tools in medicine and healthcare. Deep learning methods have achieved promising results on predictive healthcare tasks using ECG signals.

\noindent \textbf{Objective:} This paper presents a systematic review of deep learning methods for ECG data from both modeling and application perspectives.

\noindent \textbf{Methods:} We extracted papers that applied deep learning (deep neural network) models to ECG data that were published between January 1st of 2010 and February 29th of 2020 from Google Scholar, PubMed, and the Digital Bibliography \& Library Project. We then analyzed each article according to three factors: tasks, models, and data. Finally, we discuss open challenges and unsolved problems in this area.

\noindent \textbf{Results:} The total number of papers extracted was 191. Among these papers, 108 were published after 2019. Different deep learning architectures have been used in various ECG analytics tasks, such as disease detection/classification, annotation/localization, sleep staging, biometric human identification, and denoising. 

\noindent \textbf{Conclusion:} The number of works on deep learning for ECG data has grown explosively in recent years. Such works have achieved accuracy comparable to that of traditional feature-based approaches and ensembles of multiple approaches can achieve even better results. Specifically, we found that a hybrid architecture of a convolutional neural network and recurrent neural network ensemble using expert features yields the best results. 
However, there are some new challenges and problems related to interpretability, scalability, and efficiency that must be addressed. Furthermore, it is also worth investigating new applications from the perspectives of datasets and methods. 

\noindent \textbf{Significance:} This paper summarizes existing deep learning research using ECG data from multiple perspectives and highlights existing challenges and problems to identify potential future research directions. 

\end{abstract}


\maketitle

\section{Introduction}

The electrocardiogram (ECG/EKG) is one of the most commonly used non-invasive diagnostic tools for recording the physiological activities of the heart over a period of time. ECG data can aid in the diagnosis of many cardiovascular abnormalities, such as premature contractions of the atria (PAC) or ventricles (PVC), atrial fibrillation (AF), myocardial infarction (MI), and congestive heart failure (CHF). In recent years, we have witnessed the rapid development of portable ECG monitors in the medical field, such as the Holter monitor \cite{nikolic1982sudden}, and wearable devices in various healthcare areas, such as the Apple Watch. As a result, the amount of ECG data requiring analysis has grown too rapidly for human cardiologists to keep up. Therefore, analyzing ECG data automatically and accurately has become a hot research topic. Additionally, many emerging applications, such as biometric human identification and sleep staging, can be implemented based on ECG data.

\begin{figure}
\centering
\includegraphics[width=\linewidth]{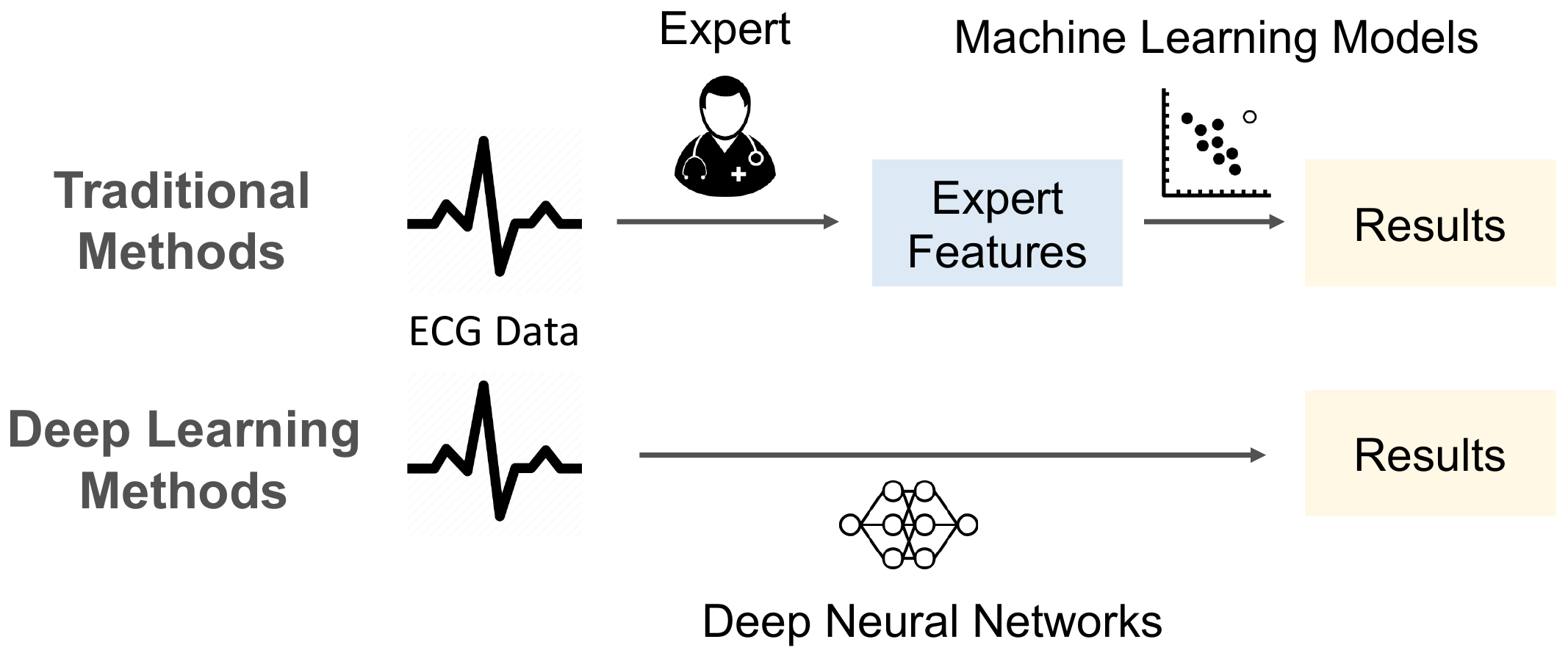}
\caption{Comparisons between traditional methods and deep learning methods.  }
\label{fig:intro}
\end{figure}

Traditionally, automatic ECG analysis has relied on diagnostic golden rules. As shown in the top of Figure \ref{fig:intro}, this is a two-stage method that requires human experts to engineer useful features based on raw ECG data, which are referred to as ``expert features'', and then deploy decision rules or other machine learning methods to generate final results. Expert features can be categorized \cite{hong2019combining} into statistical features (such as heart rate variability \cite{camm1996heart}, sample entropy \cite{alcaraz2010optimal}, and coefficients of variation and density histograms \cite{tateno2001automatic}), frequency-domain features \cite{romero2001ecg,lin2008frequency}, and time-domain features (such as the Philips 12 lead ECG Algorithm \cite{dxl}). 
In practice, expert features are automatically extracted using computer-based algorithms. However, they are still insufficient because they are limited by data quality and human expert knowledge \cite{schlapfer2017computer,shah2007errors,guglin2006common}.

Recently, deep learning methods have achieved promising results in many application areas, such as speech recognition, computer vision, and natural language processing~\cite{LeCunBH15}. The main advantage of deep learning methods is that they do not require an explicit feature extraction step using human experts, as shown in the bottom of Figure \ref{fig:intro}. Instead, feature extraction is performed automatically and implicitly by deep learning models based on their powerful data learning capabilities and flexible processing architectures. Some studies have experimentally demonstrated that deep learning features are more informative than expert features for ECG data \cite{hong2017encase,hong2019combining}. The performance of deep learning methods is also superior to that of traditional methods on many ECG analysis tasks, such as disease detection \cite{clifford2017af} and sleep staging \cite{ghassemi2018you}.

Although some papers have reviewed machine learning methods for ECG data \cite{minchole2019machine} (2019), cardiac arrhythmia detection using deep learning \cite{parvaneh2019cardiac} (2019), and deep learning methods for ECG data \cite{parvaneh2018electrocardiogram} (2018), there have no systematic reviews focusing on deep learning methods, which we consider to be promising methods for mining ECG data. Therefore, we believe it is crucial to conduct a systematic review of existing deep learning methods for ECG data from the perspectives of model architectures and application tasks. Challenges and problems related to the current research status are discussed, which should provide inspiration and insights for future work.

\section{Method}

\subsection{Search Strategy}

To conduct a comprehensive review, we searched for papers that deployed deep learning methods (deep neural networks (DNNs)) for ECG data on Google Scholar, PubMed, and the Digital Bibliography \& Library Project from January 1st of 2010 to February 29th of 2020. 

Inspired by \cite{parvaneh2019cardiac}, the following general search terms were compiled for each database: ("electrocardiogram" OR "electrocardiology" OR "electrocardiography" OR "ECG" OR "EKG" OR "arrhythmia") AND ("deep learning" OR "deep neural network" OR "deep neural networks" OR "convolutional neural network" OR "cnn" OR "recurrent neural network" OR "rnn" OR "long short term memory" OR "lstm" OR "autoencoder" OR "deep belief network" OR "dbn"). All keywords are case insensitive. 

To avoid missing papers that do not explicitly mention these keywords in their titles, we expanded our search to include all fields in each article. It should be noted that many unrelated papers mentioned some of the keywords in their introduction sections or related work sections, which resulted in a large initial set of papers.

\subsection{Study Selection}

We only included published peer-reviewed papers and excluded preprints. We focused on papers from the following journals/conferences: 
\begin{itemize}[leftmargin=5mm]
\itemsep0em
\item \textbf{Medical Information and Biomedical Engineering (MI \& BME) }: Circulation, Journal of the American College of Cardiology, Nature Medicine, Nature Biomedical Engineering, American Medical Informatics Association, Journal of American Medical Informatics Association, Journal of American Biomedical Informatics, Transactions on Biomedical Engineering, Computers in Biology and Medicine, Biomedical Signal Procession and Control, Computing in Cardiology, Physiological Measurement, IEEE Journal of Biomedical and Health Informatics, Computer Methods and Programs in Biomedicine, Journal of Electrocardiology, International Conference of the IEEE Engineering in Medicine and Biology Society. 
\item \textbf{Artificial Intelligence and Data Mining (AI \& DM)}: Nature Machine Intelligence, ACM SIGKDD International Conference on Knowledge Discovery \& Data Mining, AAAI Conference on Artificial Intelligence, International Joint Conference on Artificial Intelligence, Neural Information Processing Systems, IEEE Transactions on Pattern Analysis and Machine Intelligence, International Conference on Acoustics, Speech and Signal Processing, IEEE Transactions on Neural Networks and Learning Systems, IEEE Transactions on Knowledge and Data Engineering, IEEE Transactions on Cybernetics, Neurocomputing, Knowledge-Based System, Expert Systems with Applications, Pattern Recognition Letters. 
\item \textbf{Interdisciplinary Area}: Nature Scientific Reports, PLoS One, IEEE Access.
\end{itemize}

\begin{figure}
\centering
\includegraphics[width=\linewidth]{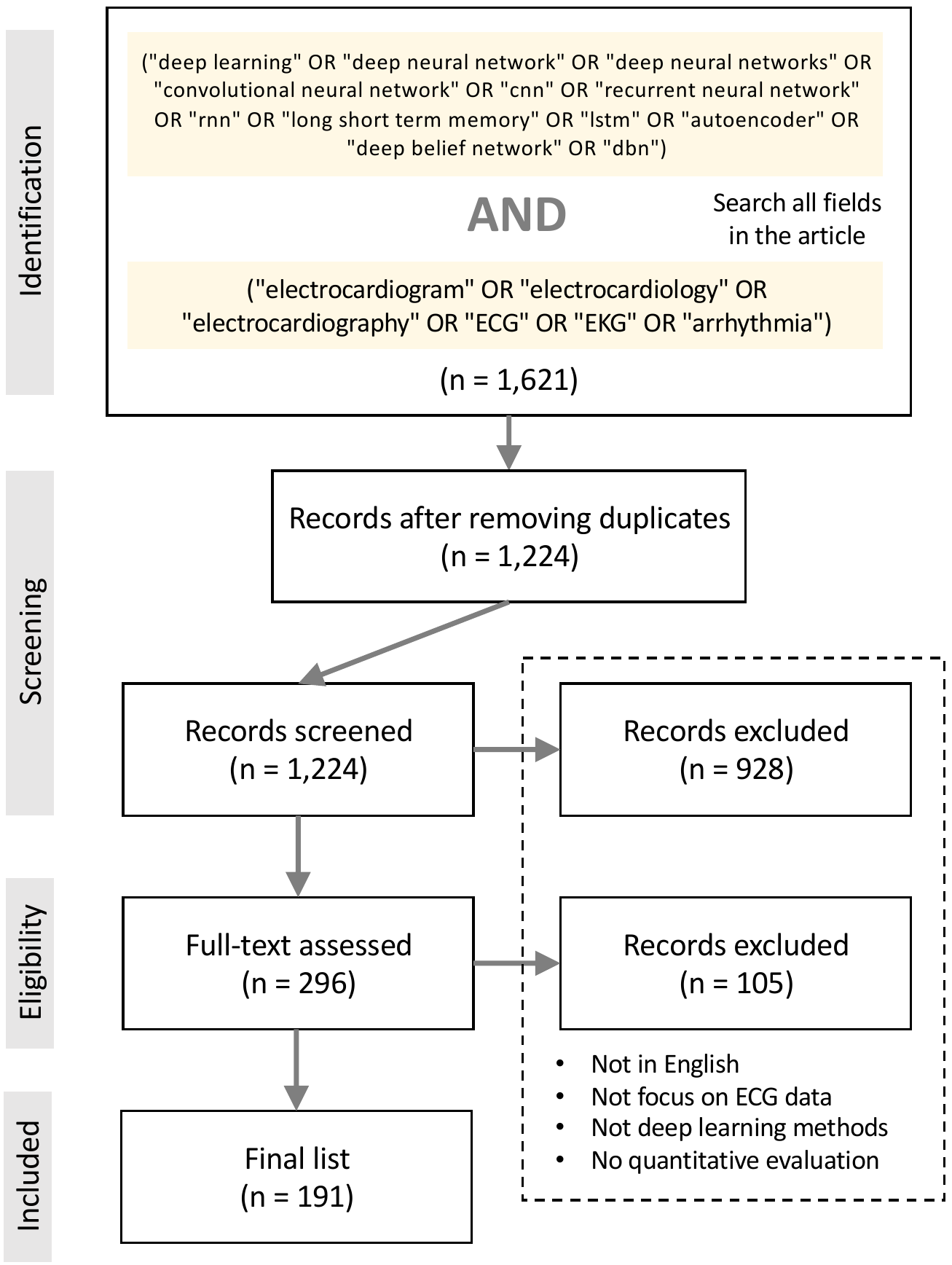}
\caption{Framework for literature searching and selection.  }
\label{fig:framework}
\end{figure}

The process of literature searching and selection is illustrated in Figure \ref{fig:framework}. It is a four-stage process consisting of identification, screening, eligibility, and inclusion. 

The number of papers collected in the identification step is 1,621. After removing duplicates, the number of papers is 1,224. Next, paper eligibility assessment was conducted in coarse-to-fine manner by two independent reviewers: one reviewer screening titles and abstracts, and another reviewer reading full texts. The exclusion criteria are 1) not in English, 2) not focusing on ECG data, 3) not using deep learning methods, and 4) no quantitative evaluations. As a result, 928 papers were excluded by screening and 105 papers were excluded by full text assessment. This left 191 papers among the initial 1,621 papers.

\subsection{Data Extraction and Analysis}

Each paper was analyzed according to the following three aspects:
\begin{itemize}[leftmargin=5mm]
\itemsep0em
\item \textbf{Task}: the targeted application tasks were (1) disease detection (e.g., specific diseases such as AF, MI, CHF, ST elevation, or general diagnostic arrhythmia), (2) annotation or localization (e.g., QRS complex annotation, P-wave annotation, localizing the origin of ventricular activation), (3) sleep staging, (4) biometric human identification, (5) denoising, and (6) others. 
\item \textbf{Model}: deep model architectures include (1) convolutional neural networks (CNNs), (2) recurrent neural networks (RNNs), (3) combinations of CNNs and RNNs (CRNNs), (4) autoencoders (AEs), (5) generative adversarial networks (GANs), (6) fully connected neural networks (FCs), and (6) others. We also determined (7) whether each paper included traditional expert features or integrated expert knowledge when constructing a deep model. 
\item \textbf{Data}: data statistics are (1) the size of the dataset, (2) number of channels (number of electrode leads), (3) duration, (4) annotations, (5) sources, (6) collection year, and (7) number of used papers.
\end{itemize}

\begin{figure*}
\centering
\includegraphics[width=\linewidth]{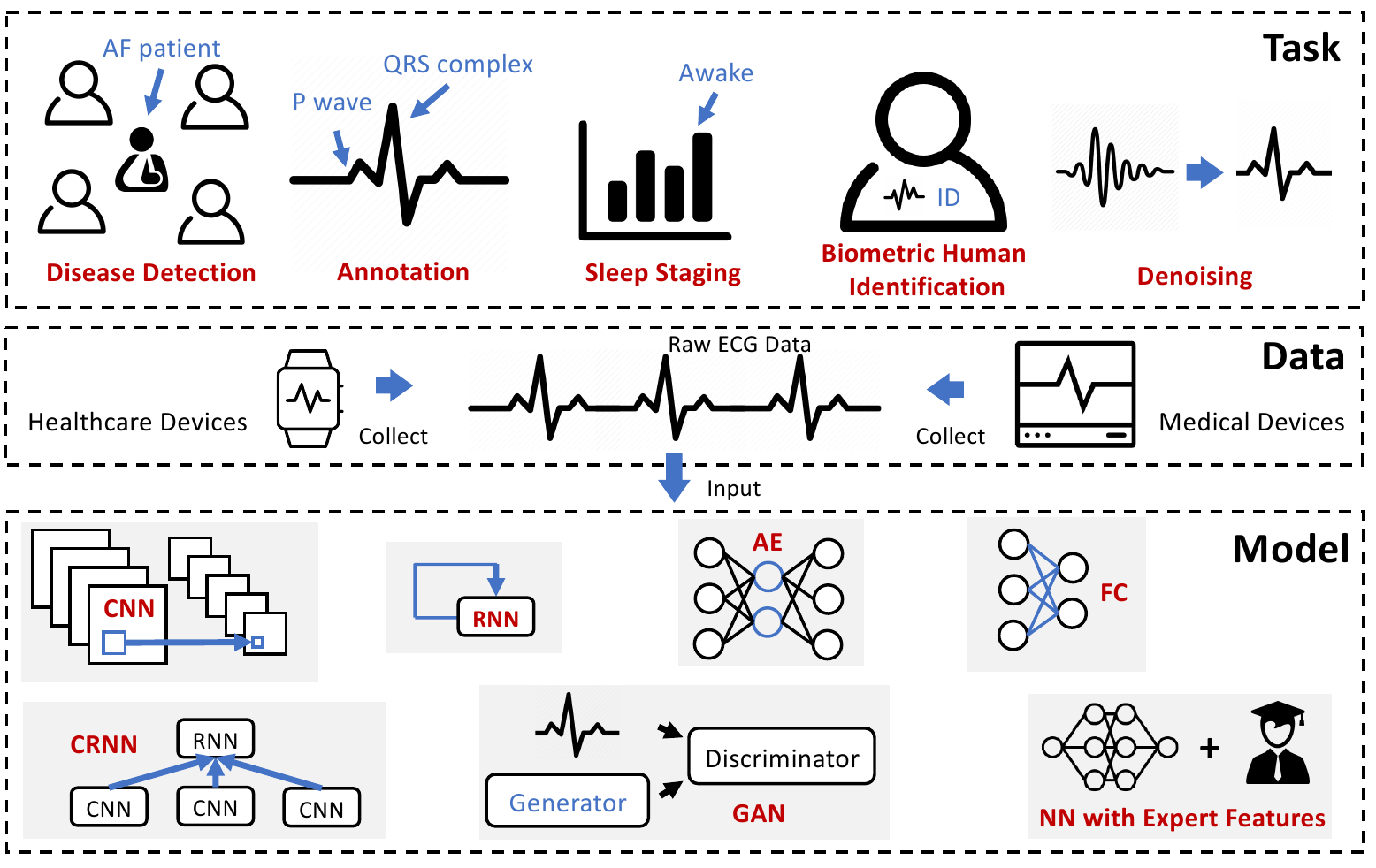}
\caption{Overview of analysis based on the aspects of tasks, models, and data.}
\label{fig:overview}
\end{figure*}

We summarized all papers from the perspectives of models and tasks, as shown in Figure \ref{fig:overview}. Finally, we discuss the challenges and problems that existing models cannot handle well. 

\section{Results}

\begin{figure}
\centering
\includegraphics[width=\linewidth]{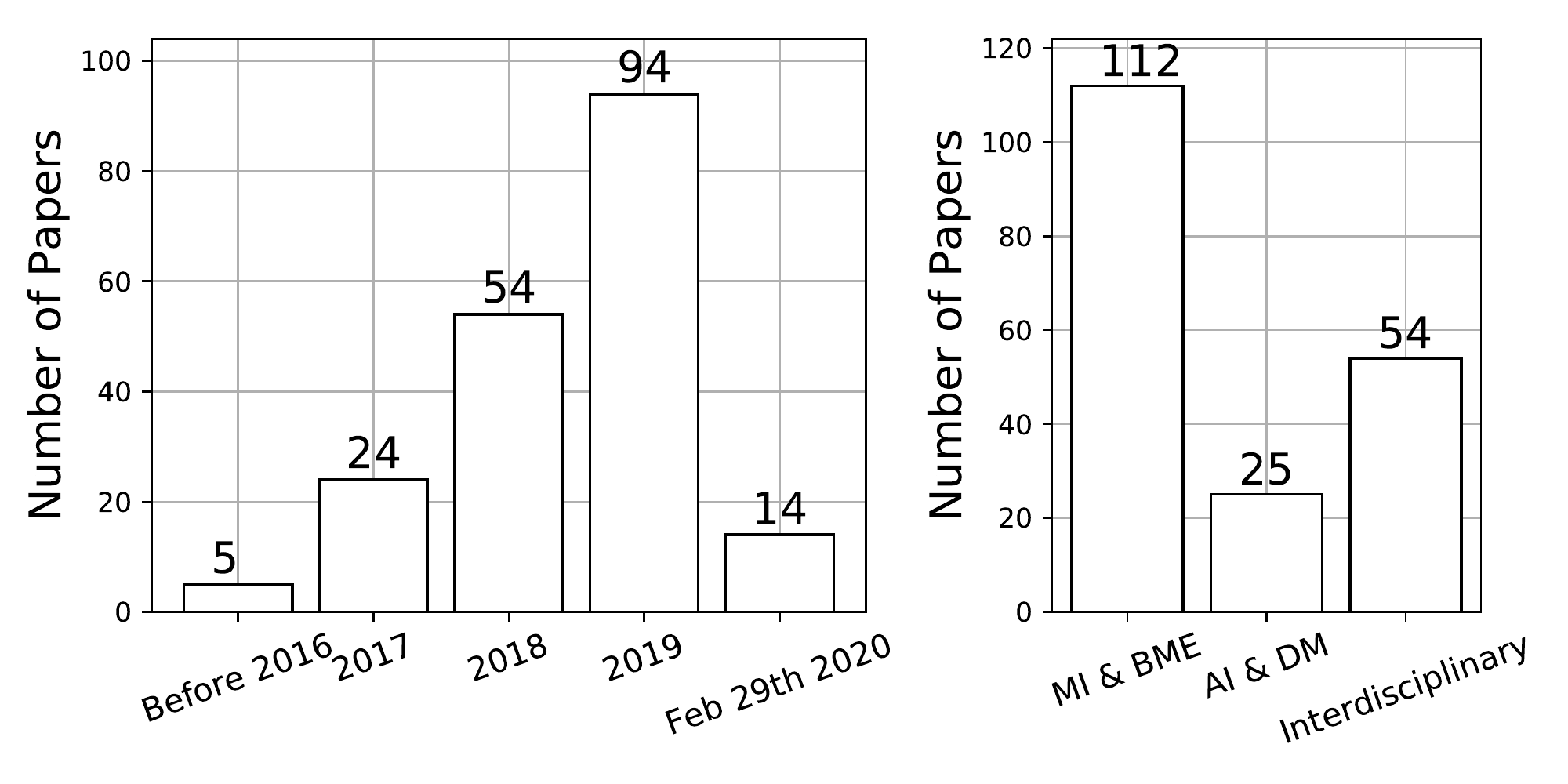}
\caption{General statistics for all papers.}
\label{fig:stat}
\end{figure}

We included 191 papers in our survey and analyzed them based on the aspects of tasks, models, and data. An overview of our analysis is presented in Figure \ref{fig:overview}. General statistics for these papers are presented in Figure \ref{fig:stat}. Among the included papers, 108 (approximately 57\%) were published after 2019, 112 were published by the medical information and biomedical engineering community, and only 25 (approximately 13\%) were published by the artificial intelligence and data mining community. 
We provide a detailed summary of each paper on our GitHub at \url{https://github.com/hsd1503/DL-ECG-Review}. 

\begin{table*}
\centering
\resizebox{1.0\linewidth}{!}{
\begin{tabular}{p{2cm}p{6cm}p{1.8cm}p{1.5cm}p{2cm}p{1.5cm}p{1.5cm}}
\toprule
& Disease Detection & Annotation or Localization & Sleep Staging & Biometric Identification & Denoising & Others \\
\midrule
CNN & \cite{kiranyaz2015real,kiranyaz2015convolutional,yin2016ecg,liu2017real,acharya2017deep,hong2017encase,xiong2017robust,andreotti2017comparing,xia2017atrial,rubin2017densely,limam2017atrial,ghiasi2017atrial,kiranyaz2017personalized,pourbabaee2017deep,xiao2018deep,xia2018detecting,yildirim2018arrhythmia,attia2018electrocardiographic,gadaleta2018deep,golrizkhatami2018ecg,li2018combining,zhai2018automated,li2018patient,parvaneh2018analyzing,xiong2018ecg,plesinger2018parallel,sodmann2018convolutional,kamaleswaran2018robust,strodthoff2018detecting,nguyen2018deep,xiao2018monitoring,rubin2018densely,cai2019accurate,jiang2019novel,sellami2019robust,dang2019novel,he2019automatic,wolk2019early,huang2019ecg,he2019simultaneous,xia2019novel,sun2019morphological,li2019automated,li2019automated_repeat,he2019real,cao2019atrial,park2019preprocessing,li2019classification,shen2019ambulatory,attia2019screening,hannun2019cardiologist,li2019ventricular,hong2019combining,niu2019inter,vullings2019fetal,zhao2019pvc,selvalingam2019developing,zhou2019beatgan,baloglu2019classification,brisk2019deep,lai2019automatic,wu2019deep,hao2019spectro,tadesse2019cardiovascular,jalali2019atrial,noseworthy2020assessing,van2020transfer,hao2020multi,han2020ml} & \cite{yang2017localization,camps2018deep,zhong2018deep,chen2018region,yu2018qrs,lee2019qrs,habib2019impact,jimenez2019u,tison2019automated,smith2019deep,jia2019electrocardiogram,viktor2020deep} & \cite{malik2018sleep,urtnasan2018multiclass,erdenebayar2019deep} & \cite{zhang2017heartid,zhao2018ecg,hammad2018multimodal,chen2018finger,chu2019ecg,labati2019deep,zhang2019ecg,hong2019ecg,li2020toward} & \cite{zhao2019noise,chiang2019noise,fotiadou2019deep,fotiadou2020end} & \cite{santamaria2018using,attia2018noninvasive,miller2019comparison,baalman2019deep,selvalingam2019developing,attia2019age,ye2019ecg,shaker2020generalization} \\
RNN & \cite{maknickas2017atrial,schwab2017beat,yildirim2018novel,chang2018af,teijeiro2018abductive,rajan2018generative,liu2018classification,zhou2018premature,chauhan2019ecg,dang2019novel,he2019automatic,wang2019automatic,li2019interpretability,golany2019pgans,zhang2019localization,xu2019parallel,saadatnejad2019lstm} & \cite{attin2017annotating,peimankar2019ensemble,gyawali2019sequential} & \cite{wei2019multi,erdenebayar2019deep} & N.A. & \cite{qiu2017elimination} & \cite{harada2019biosignal} \\
CRNN & \cite{zihlmann2017convolutional,warrick2017cardiac,tison2017cardiovascular,ballinger2018deepheart,faust2018automated,tan2018application,oh2018automated,rajan2018generalization,shashikumar2018detection,warrick2018ensembling,xie2018bidirectional,andersen2019deep,liu2019lstm,mousavi2019inter,wang2019deep,wang2019ecg,hong2019mina,zhou2019k,picon2019mixed,jia2019detection,wang2019pay,yildirim2019new,li2019dual,xie2019feature,al2019dense,zhao2019deep,liu2019mfb,mousavi2020single,li2020heartbeat} & \cite{yuen2019inter} & N.A. & \cite{lynn2019deep} & N.A. & \cite{elola2018deep,rastgoo2019automatic,xu2018raim,goto2019artificial,zhu2019electrocardiogram,porumb2020precision} \\
AE & \cite{farhadi2018classification,xia2018automatic,wang2019automatic,zhang2019automated,yildirim2019new} & \cite{gyawali2019sequential,gyawali2017automatic} & \cite{li2018method} & N.A. & \cite{chiang2019noise,xiong2016stacked} & \cite{gogna2016semi,yin2017recognition,kuznetsov2020electrocardiogram} \\
GAN & \cite{wang2019ecg,golany2019pgans,zhou2019beatgan} & N.A. & N.A. & N.A. & \cite{wang2019adversarial} & \cite{lee2019synthesis,zhu2019electrocardiogram,ye2019ecg,harada2019biosignal,shaker2020generalization} \\
NN with Expert Features & \cite{hong2017encase,andreotti2017comparing,xia2017atrial,maknickas2017atrial,ghiasi2017atrial,golrizkhatami2018ecg,huang2019ecg,li2019interpretability,teijeiro2018abductive,kamaleswaran2018robust,hong2019combining,saadatnejad2019lstm,yang2020detection,li2019dual,lai2019automatic,hao2019spectro,plesinger2018parallel,sodmann2018convolutional,parvaneh2018analyzing} & N.A. & N.A. & \cite{hammad2018multimodal} & \cite{zhao2019noise} & N.A. \\
FC \& Others & \cite{xu2018towards,xu2019vector,mathews2018novel,oh2019automated,maidens2018artificial,kimura201810,feng2019probabilistic,mukherjee2019detection,yang2020detection} &N.A. & \cite{erdenebayar2019deep} & \cite{lee2019personal} & N.A. & \cite{zhang2017ecg,kwon2019artificial} \\
\bottomrule
\end{tabular}
}
\caption{Summary of papers from the perspectives of models and tasks.}
\label{tb:stat}
\end{table*}

\subsection{Task}

\subsubsection{Disease Detection}
The goal of developing a deep learning model for disease detection is to map input ECG data to output disease targets through multiple layers of neural networks \cite{hong2020cardiolearn}. For example, the detection of cardiac arrhythmias (e.g., atrial flutter, supraventricular tachyarrhythmia, and ventricular trigeminy) \cite{hannun2019cardiologist} is one of the most common tasks for deep learning models based on ECG signals. AF detection can be regarded as a special case of cardiac arrhythmia detection, where all non-AF rhythms are grouped together \cite{teijeiro2018abductive, zhou2019k}. Additionally, a deep learning technique was introduced to monitor ST changes in ECG data \cite{park2019preprocessing}. In \cite{liu2017real} and \cite{wang2019deep}, convolutional neural networks were applied to automate the detection of MI and CHF, respectively. 

From the perspective of the volume of classification results, multiple modalities (e.g., ECG, transthoracic echocardiogram \cite{attia2019screening}) can be employed for binary classification tasks (e.g., patients with paroxysmal AF or healthy patients \cite{pourbabaee2017deep}), multi-class classification tasks (e.g., detection of acute cognitive stress \cite{he2019real} and decompensation of patient detection \cite{xu2018raim}), and multi-task classification tasks (e.g., detection of prevalent hypertension, sleep apnea, and diabetes \cite{tison2017cardiovascular}). 

\subsubsection{Localization and Annotation}
The localization and annotation of specific waves in ECG signals are of great importance for cardiologists attempting to diagnose cardiac diseases, such as AF. For instance, AEs~\cite{gyawali2017automatic} and RNNs~\cite{gyawali2019sequential} have been employed to localize the exit of ventricular tachycardia in 12-lead ECG data automatically. Some studies have focused on localizing of the origins of premature ventricular issues~\cite{yang2017localization} and MI~\cite{zhang2019localization}.

Most studies that have applied deep learning methods to ECG annotation have focused on using deep learning methods for annotating the fetal QRS complex (detecting Q-waves, R-waves, and S-waves, and calculating heart rate), which is critical for identifying various arrhythmias, in the MIT-BIH arrhythmia dataset \cite{lee2019qrs}. Some studies have used the QT database of PhysioNet~\cite{goldberger2000physiobank} to explore other methods for annotating ECG waves, including P-wave \cite{peimankar2019ensemble} and T-wave annotation \cite{attin2017annotating}.
Another annotated 12-lead ECG dataset named Lobachevsky University database (LUDB) is provided by \cite{kalyakulina2018lu}.

\subsubsection{Sleep Staging}
Understanding sleep patterns is critical for improving healthcare because sleep is a key aspect of our wellbeing. Sleep disorders can lead to catastrophes in personal medicine or public health \cite{malik2018sleep}. In \cite{li2018method}, a sparse AE (SAE) and hidden Markov model (HMM) were combined to detect obstructive sleep apnea (OSA) using the PhysioNet challenge 2000 dataset. A CNN has also been used to detect OSA \cite{malik2018sleep}.

In addition to OSA detection, CNN-based deep learning architectures have also been employed for the multi-class classification of OSA with hypopnea \cite{urtnasan2018multiclass}, which is the most common sleep-related breathing disorder, based on single-lead ECG data. Some studies have focused on sleep stage identification outside of OSA detection. For example, a long short-term memory (LSTM) network was employed in \cite{wei2019multi} to analyze a multi-channel physiological signal dataset (electroencephalogram, electrooculogram, and electromyogram signals), which was collected by the Sleep Disorder Diagnosis Center of Xijing Hospital at the Fourth Military Medical University.

\subsubsection{Human Identification}
Based on the rapid development of information technology, body sensor networks are reshaping people's daily lives, particularly in the context of smart health applications. Biometric-based human identification is a promising technology for automatic and accurate individual recognition based on various body sensor data, including heart rate, temperature, and activity.~\citeauthor{zhang2017heartid} constructed a CNN-based biometric human identification system, evaluated their system using eight datasets from PhysioNet (CEBSDB, WECG, FANTASIA, NSRDB, STDB, MITDB, AFDB, and VFDB), and achieved state-of-the-art performance using a CNN. Similarly, the PTB diagnostic dataset (549 15 channel ECG records from 290 subjects) \cite{labati2019deep} and CYBHi dataset (65 subjects between 21.64 and 40.56 y old) \cite{hammad2018multimodal} were also used to evaluate CNN-based biometric human identification methods.

Some studies have combined CNNs with other approaches.~\citeauthor{zhao2018ecg} proposed a novel ECG biometric authentication system that incorporates the generalized S-transformation and CNN techniques. A secure multimodal biometric system using a CNN and Q-Gaussian multi-support vector machine (SVM) based on a different levels of fusion was developed in~\cite{hammad2018multimodal}. 

Additionally, residual networks have been employed to develop ECG biometric authentication methods that can improve generalization for ECG signals sampled in the different environments for similar tasks \cite{chu2019ecg}. In \cite{lee2019personal}, features were extracted using a principal component analysis network (PCANet). Next, a robust Eigen ECG network was applied to the time-frequency representations of ECGs for personal identification. A bidirectional gated recurrent unit (GRU)-based method was also proposed for human identification based on ECG data \cite{lynn2019deep}.

\subsubsection{Denoising}
The ECG signal acquisition process is often accompanied by a large amount of noise, which negatively affects the accuracy of diagnosis, particularly in telemedicine environments. In \cite{fotiadou2020end}, an encoder-decoder CNN, which is a widely used deep learning technique, was employed for ECG denoising. Similarly,~\citeauthor{xiong2016stacked} and~\citeauthor{chiang2019noise} proposed fully-convolutional-network-based denoising AE (DAE) methods for ECG signal denoising. In \cite{lee2018bidirectional}, a bidirectional recurrent DAE was used to perform photoplethysmography feature accentuation for pulse waveform analysis.

Additionally, GANs have been employed to accumulate knowledge regarding the distribution of ECG noise continuously through a minimax game between a generator and discriminator, where the quality of denoised signals was evaluated versus an SVM algorithm~\cite{wang2019adversarial}. \citeauthor{zhao2019noise} addressed noise by developing a noise rejection method based on a combination of a modified frequency slice wavelet transform (WT) and CNN.

\subsubsection{Others}
The CNNs used in other types of studies focused on emotion detection \cite{santamaria2018using}, drug assessment \cite{attia2018noninvasive}, data compression \cite{zhang2017ecg}, transfer learning in ECG classification from human data to horse data \cite{van2020transfer}, etc. Combinations of CNNs and LSTM networks have been used to learn long-term dependencies (e.g., classifying pulsed rhythm or pulseless electrical activity \cite{elola2018deep}, detecting hypoglycemic events in healthy individuals \cite{porumb2020precision}, predicting the need for urgent revascularization \cite{goto2019artificial}, and classifying driver stress levels \cite{rastgoo2019automatic}). A combination of a CNN and HMM was used to segment ECG data into standard component waveforms and intervals \cite{tison2019automated}.
 
Additionally, AEs \cite{gogna2016semi} have been used for the reconstruction and analysis of biomedical signals. GANs have been employed for ECG generation \cite{zhu2019electrocardiogram} and synthesis \cite{lee2019synthesis}, as well as anomaly beat detection \cite{zhou2019beatgan}. Some studies have used DNNs to process raw ECG waveform data for future risk prediction (e.g., assessing the risk of future cardiac disease \cite{miller2019comparison}, sudden cardiac arrest risk prediction \cite{selvalingam2019developing}, mortality prediction \cite{raghunath2019deep}, and physiological measures for health prediction \cite{attia2019age}).

\subsection{Models}

\subsubsection{CNN}\label{sec:CNN}

CNNs represent a class of DNNs that are widely applied for image classification, natural language processing, and signal analysis. Such networks can automatically extract hierarchical patterns in data using stacked trainable small filters or kernels, meaning they require relatively little preprocessing compared to handcrafted features. A typical CNN is composed of several convolutional layers followed by a batch normalization layer, nonlinear activation layer, dropout layer, pooling layer, and classification layer (e.g., fully connected layer), as discussed in~\cite{hannun2019cardiologist}. In some works, SVMs, boosting classifier trees, and RNNs have been used as alternative fully connected layers to summarize global features in CNNs. CNNs can achieve superior performance and fast computation based on shared-weight architectures and parallelization. 

Two types of CNN are commonly used for ECG classification, namely the 1D CNN and 2D CNN. A 1D CNN operates by applying kernels along the temporal dimension of raw ECG data, whereas a 2D CNN typically operates on transformed ECG data, such as distance distribution matrices based on entropy calculations~\cite{li2018combining}, gray-level co-occurrence matrices~\cite{sun2019morphological}, or combined features, such as morphologies, RR intervals, and beat-to-beat correlations~\cite{li2019automated}. However, there is some controversy regarding the use of multi-head ECG or time-frequency ECG spectrograms extracted using the WT, fast Fourier transform (FFT), or short-term Fourier transform. Some works, such as~\cite{huang2019ecg}, have directly applied 2D-CNNs to such spectrograms, but this introduces issues related to varying frequency resolutions, meaning most signal characteristics are reflected by intra-component patterns, rather than inter-component behaviors. To overcome this issue, a shared 1D CNN was adopted in~\cite{kiranyaz2015real, liu2017real}. Similarly, a multi-scale 1D CNN was adopted in~\cite{zhang2017heartid} for biometric human identification.

Additionally, a 2D CNN can be applied to a one-head ECG signal, which is treated as an image. In such cases, a pre-trained ResNet, DenseNet, or Inception-Net trained on an ImageNet dataset~\cite{deng2009imagenet} can be fine tuned on an ECG dataset to perform heartbeat disease detection~\cite{wolk2019early} and analyze ST changes~\cite{wang2019ecg}. Particularly for localization tasks, such as QRS detection, a fully convolutional network with a large kernel size can be applied to an ECG image to reduce that image's height to one while maintaining an output length equal to the input length to derive a QRS window label~\cite{lee2019qrs}.

Some advanced techniques, such as an atrous spatial pyramid pooling module, can be used to exploit multi-scale features in ECG data. Additionally, active learning~\cite{xia2019novel} and data augmentation~\cite{wang2019ecg, zhou2019k} can be incorporated into CNN frameworks to handle imbalance problems and further improve accuracy.

\subsubsection{RNN}\label{sec:RNN}

An RNN is a type of neural network designed to model sequential data, such as time series, event sequences, and natural language. In an RNN, the output from the previous step is used as the input for the current step. By iteratively updating hidden states and memory, an RNN is capable of remembering information in sequential order. 

In particular, for ECG data, RNNs are a logical choice for both capturing temporal dependencies and handling inputs of various lengths. GRU/LSTM and bidirectional-LSTM (BiLSTM) are commonly used RNN variants that handle a critical problem called the vanishing gradient, which negatively affects classical RNNs. Two small LSTM networks were employed in~\cite{saadatnejad2019lstm} to combine raw ECG features and WT features for continuous real-time execution on wearable devices. In~\cite{li2019interpretability}, an attention mechanism was incorporated into a BiLSTM network to improve performance and interpretability by visualizing attention weights. 

\subsubsection{CRNN}\label{sec:CRNN}
A CRNN, as the name suggests, combines CNN and RNN modules. This is a preferred architecture for handling long ECG signals with varied sequence lengths and multi-channel inputs. A 1D CNN~\cite{he2019automatic} or 2D CNN~\cite{shashikumar2018detection} is used to extract local features from an ECG sequence. An RNN then summarizes local features along the time dimension to generate global features. 

DeepHeart~\cite{ballinger2018deepheart} follows the CRNN framework to perform cardiovascular risk prediction for heart rate sequences extracted from ECG data and utilizes an AE model (see Sec.~\ref{sec:AE}) to initialize model weights to achieve enhanced performance. To provide interpretable diagnosis, MINA~\cite{hong2019mina} incorporates a CRNN with a multi-level attention mechanism with beat-level, rhythm-level, and frequency-level expert features based on medical domain knowledge.

\subsubsection{AE}\label{sec:AE}
An AE is a type of neural network composed of an encoder module and decoder module for learning embeddings in an unsupervised manner. The goal of an AE is to learn a reduced dimensional representation using the encoder module while the decoder attempts to reconstruct an original input from this reduced representation. There are three commonly used variants of the AE, namely the DAE, SAE, and contractive AE (CAE). DAEs take partially corrupted inputs and are trained to recover original undistorted inputs. SAEs and CAEs utilize different regularization methods, such as KL-divergence and the Frobenius norm of the Jacobian matrix, to learn robust representations.

Stacked DAEs~\cite{xia2018automatic}, SAEs~\cite{yin2017recognition, zhang2019automated}, and CAEs~\cite{xiong2016stacked} are widely used for ECG denoising purposes because ECG signals are often contaminated by various types of noise, such as baseline wandering, electrode contact noise, and motion artifacts, which may lead to inaccurate interpretations. In practice, the effectiveness of an AE is determined by the choice of encoder and decoder modules. As introduced in Sections~\ref{sec:CNN},~\ref{sec:RNN}, and~\ref{sec:CRNN}, CNNs, RNNs, and CRNNs are typically combined to serve as encoder and decoder modules. The authors of~\cite{chiang2019noise} utilized fully convolutional networks as an encoder and decoder while~\cite{laguna1997database} used BiLSTM. To improve classification performance, the authors of~\cite{gogna2016semi, rajan2018generative, yildirim2019new} simultaneously conducted the reconstruction and classification procedures. For examples, the authors of~\cite{yildirim2019new} utilized a 1D CNN AE to first compress large ECG signals with minimum loss, and then fed the compressed signals into an LSTM network to recognize arrhythmias automatically.

\subsubsection{GAN}\label{sec:GAN}
A GAN is a class of neural network framework that was invented by Goodfellow et al.~\cite{goodfellow2014generative}. This type of model consists of two sub-models: a generative model $G$ that captures the data distribution of a training dataset in a latent representation and a discriminative model $D$ that determines the probability that a sample produced by the generator comes from the true data distribution. These two models are trained iteratively to conduct a minimax game.

GAN application has increased rapidly, particularly in areas such as image generation~\cite{brock2018large} and language generation~\cite{lin2017adversarial}. Recently, GANs have been applied to tackle the data imbalance challenge in ECG data. For example, the authors of~\cite{wang2019ecg} proposed an abnormality detection model for ECG signals based on a CRNN framework and used a GAN composed of multiple 1D CNNs to perform data augmentation. Their model achieved excellent performance for class-imbalanced datasets. The authors of~\cite{wolk2019early} utilized a GAN to perform denoising of ECG data and the authors of~\cite{zhu2019electrocardiogram} proposed a GAN composed of a BiLSTM network (generator) and CNN (discriminator) to generate synthetic ECG data to train a deep learning model. Similarly, a model called PGAN~\cite{golany2019pgans} was proposed to perform personalized ECG
classification, where subject-specific labeled data is sparse. Specifically, a GAN was optimized using a specialized loss function and trained to generate personalized synthetic ECG signals for different arrhythmias.

\subsubsection{NN with Expert Features}

ENCASE~\cite{hong2017encase} suggests that expert features can be divided into three categories: 1) statistical features (e.g., count, mean, maximum, and minimum), 2) signal processing features that transform ECG data from the time domain into the frequency domain (e.g., FFT, WT, and Shannon entropy), and 3) medical features based on medical domain knowledge (e.g., features based on P-, Q-, R-, S- and T-waves, sample entropy, and the coefficient of variation and density histograms). 

All of these methods can benefit significantly from expert features, although additional effort is required to extract such features compared to raw morphological features. Ensemble methods~\cite{hong2017encase, huang2019ecg, xia2017atrial,plesinger2018parallel,sodmann2018convolutional,parvaneh2018analyzing} combine expert features with raw morphological features and incorporate various tree models with DNNs to achieved better results compared to individual models.

\subsubsection{FC \& Others}
Some works have relied on FCs for disease detection and classification~\cite{feng2019probabilistic, kimura201810}, particularly for extremely short ECG sequences (e.g., 10 RR intervals)~\cite{kimura201810}. Stacked restricted Boltzmann machines, which are a type of generative stochastic DNN, are also used to process raw ECG data in combination with beat alignment processing to classify heartbeat types~\cite{xu2018towards}.

Other works have borrowed ideas from the computer vision field. For example, U-net~\cite{ronneberger2015u}, which is based on a fully convolutional network and consists of a contracting path and expansive path, is widely used for image segmentation tasks. The authors of~\cite{oh2019automated} proposed a modified U-net to handle various lengths of ECG sequences for classification and R-peak detection. PCANet~\cite{chan2015pcanet}, which is an image classification model based on cascaded principal component analysis, binary hashing, and block-wise histograms, was modified for biometric human identification tasks in a non-stationary ECG noise environment~\cite{lee2019personal}. However, the effectiveness of this method must still be evaluated according to the state-of-the-art baselines discussed above.

\begin{figure}
\centering
\includegraphics[width=\linewidth]{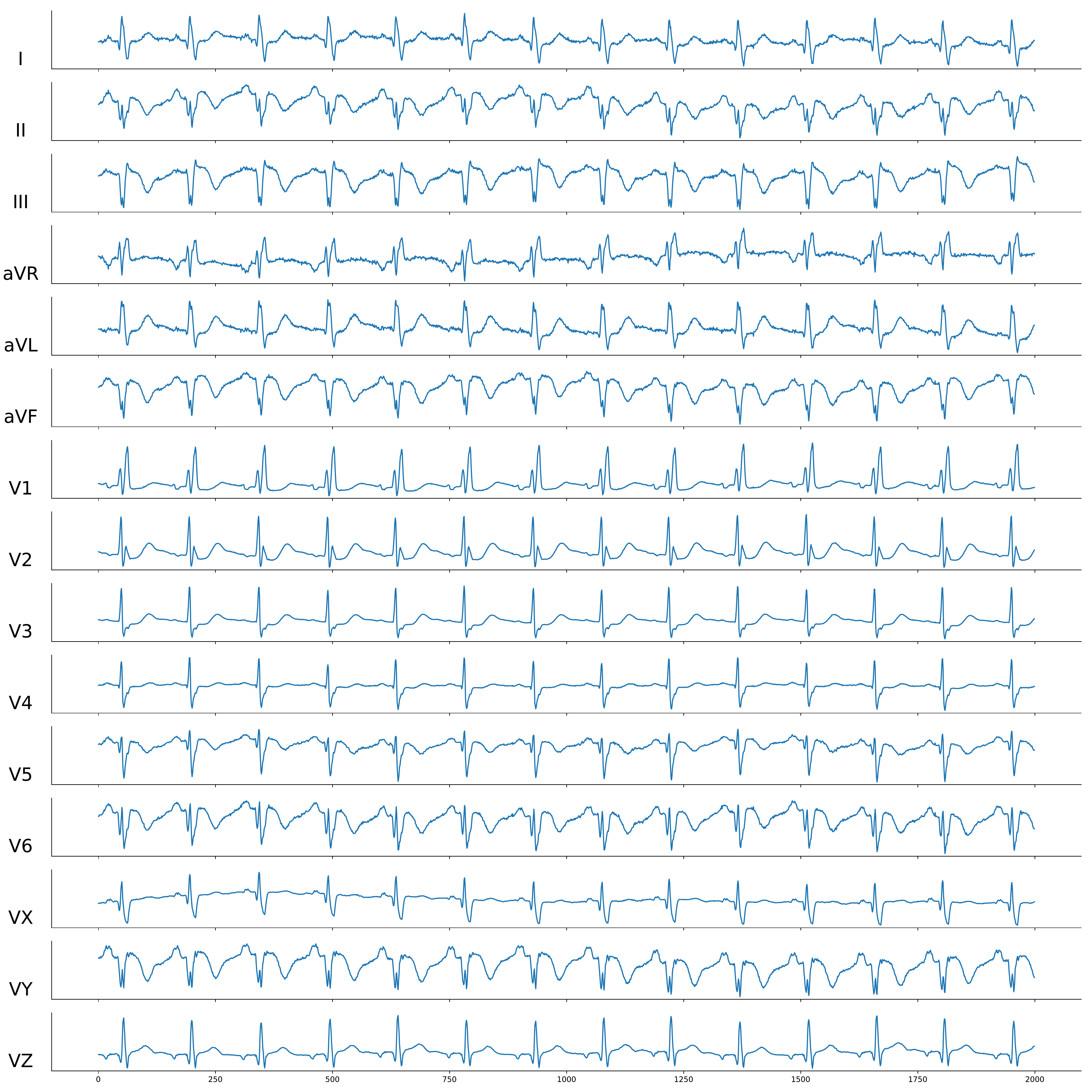}
\caption{Example 15 lead ECG data collected from medical devices. The figure presents 10 s of 15 lead ECG data from an MI patient in the PTB Diagnostic ECG Database.  }
\label{fig:data_12}
\end{figure}

\begin{figure}
\centering
\includegraphics[width=\linewidth]{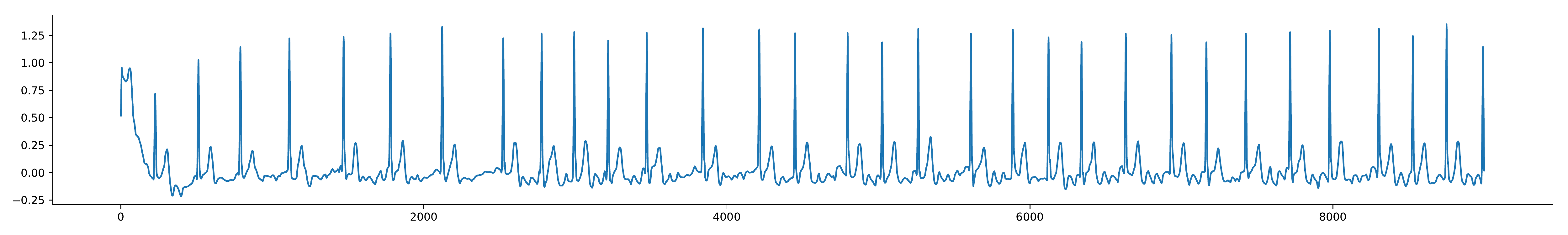}
\caption{Example single-lead ECG data collected from healthcare devices. The figure presents 30 s of single-lead ECG data from an AF patient in the PhysioNet Computing in Cardiology Challenge 2017 dataset.  }
\label{fig:data_single}
\end{figure}

\begin{table*}[]
\scriptsize
\begin{tabular}{p{2.5cm}|p{1.3cm}p{0.5cm}p{1.3cm}p{5.5cm}p{2cm}p{0.5cm}p{0.5cm}}
\toprule
Database                                         & Records               & Leads & Duration        & Annotations                                                                                                                                                                                                                                & Source                                             & Year      & Papers \\ \midrule
MIT-BIH Arrhythmia Database  \footnotemark                    & 47                    & 2     & 30 minutes      & Beat Level, rhythm level. (annotations are keeping updating, please refer to https://archive.physionet.org/physiobank/annotations.shtml)                                                                                                   & Boston's Beth Israel Hospital                      & 1975-1979 & 54     \\ \hline
PhysioNet Computing in Cardiology Challenge 2017 \footnotemark & 8528 train, 3658 test & 1     & 30 seconds      & Rhythm level: normal, AF, other, noise                                                                                                                                                                                                     & AliveCor healthcare device                         & 2017      & 33     \\\hline
PTB Diagnostic ECG Database  \footnotemark                    & 549                   & 15    & Several minutes & Rhythm level: Myocardial infarction, Cardiomyopathy/Heart failure, Bundle branch block, Dysrhythmia, Myocardial hypertrophy, Valvular heart disease, Myocarditis, Miscellaneous, Healthy controls                                          & National Metrology Institute of Germany            & 1995      & 16     \\\hline
MIT-BIH Atrial Fibrillation Database  \footnotemark           & 25                    & 2     & 10 hours        & Rhythm level: AF, atrial flutter (AFL), AV junctional rhythm, and others                                                                                                                                                                   & Boston's Beth Israel Hospital                      & 1983      & 8      \\\hline
2018 China Physiological Signal Challenge  \footnotemark      & 6877 train, 2954 test & 12    & 15 seconds      & Rhythm level: AF, I-AVB, LBBB, RBBB, PAC, PVC, STD, STE                                                                                                                                                                                    & 11 hospitals                                       & 2018      & 7      \\\hline
QT Database \footnotemark                                     & 105                   & 2     & 15 minutes      & Onset, peak, and end markers for P, QRS, T, and U waves                                                                                                                                                                                    & Compiled from several existing databases           & 1997      & 6      \\\hline
MIT-BIH Normal Sinus Rhythm Database \footnotemark            & 18                    & 2     & 24 hours        & Beat Level: normal                                                                                                                                                                                                                         & Boston's Beth Israel Hospital                      & N.A.      & 5      \\\hline
St Petersburg INCART 12-lead Arrhythmia Database \footnotemark & 75                    & 12    & 30 minutes      & Rhythm level: Acute MI, Transient ischemic attack (angina pectoris), Prior MI, Coronary artery disease with hypertension, Sinus node dysfunction, Supraventricular ectopy, Atrial fibrillation or SVTA, WPW, AV block, Bundle branch block & St. Petersburg Institute of Cardiological Technics & 2003      & 3      \\\hline
MIT-BIH Malignant Ventricular Ectopy Database \footnotemark   & 22                    & 2     & 30 minutes      & ventricular tachycardia, ventricular flutter, and ventricular fibrillation                                                                                                                                                                 & Compiled from two separate databases               & 1986      & 3      \\\hline
CU Ventricular Tachyarrhythmia Database \footnotemark         & 35                    & 1     & 8 minutes       & ventricular tachycardia, ventricular flutter, and ventricular fibrillation                                                                                                                                                                 & Creighton University Cardiac Center                & 1986      & 3     \\
\bottomrule
\end{tabular}
\caption{Summary of databases.  }
\label{tb:data}
\end{table*}

\subsection{Data}

Most of the works identified in this review (150 out of 191) used open-source datasets, which makes it easier to perform follow-up works and reproduction. 
A summary of the 10 most frequently used open-source databases is provided in Table \ref{tb:data}. We list the following parameters for each database, which can be used to classify the databases into further subtypes: 
\begin{itemize}[leftmargin=5mm]
\itemsep0em
\item Sources. Most databases were collected from medical devices, but a few were collected from healthcare devices. The biggest difference between such devices is that medical device data typically have more leads than healthcare device data, meaning medical device data are more informative. However, medical device data are also more difficult to collect. Healthcare ECG monitor devices, such as smart hardware and wristbands, are becoming increasingly common. 
\item Number of leads. A standard 12 lead ECG (or even 15-lead ECG) system can observe more abnormalities than a single-lead ECG (similar to lead I in a 12 lead ECG). For example, posterior wall MI can only be detected by chest leads (V1 to V4) and no abnormalities will be detected by a single lead. Examples of 15 lead ECG data from medical devices and single lead ECG data from healthcare devices are presented in Figures \ref{fig:data_12} and \ref{fig:data_single}, respectively.
\item Duration. Short-term ECG data (less than several minutes) and long-term ECG data can complement each other. Short-term ECG data is cheaper and easier to collect. Many cardiac diseases can be detected based on short-term ECG data, so such data represent the primary diagnostic tool in outpatient departments. However, long-term ECG can help to detect diseases with intermittent symptoms, such as paroxysmal ventricular fibrillation (VF) and AF. 
\item Annotations. Annotations include ECG measurement annotations (onset, peak, and end markers for P-, QRS-, T-, and U-waves), beat-level annotations (PAC, PVC, etc.), and rhythm-level annotations (covers both beat-level annotations and other diseases such as AF and VF). Annotation requires huge effort by medical experts. 
\end{itemize}

\footnotetext[1]{\url{https://physionet.org/content/mitdb/1.0.0/}}
\footnotetext[2]{\url{https://www.physionet.org/content/challenge-2017/1.0.0/}}
\footnotetext[3]{\url{https://physionet.org/content/ptbdb/1.0.0/}}
\footnotetext[4]{\url{https://physionet.org/content/afdb/1.0.0/}}
\footnotetext[5]{\url{http://2018.icbeb.org/Challenge.html}}
\footnotetext[6]{\url{https://physionet.org/content/qtdb/1.0.0/}}
\footnotetext[7]{\url{https://physionet.org/content/nsrdb/1.0.0/}}
\footnotetext[8]{\url{https://physionet.org/content/incartdb/1.0.0/}}
\footnotetext[9]{\url{https://physionet.org/content/vfdb/1.0.0/}}
\footnotetext[10]{\url{https://physionet.org/content/cudb/1.0.0/}}

The following databases were used by many of the selected papers: 

\begin{itemize}[leftmargin=5mm]
\itemsep0em
\item The MIT-BIH Arrhythmia Database \cite{moody2001impact} (54 papers) consists of 48 half-hour ECG records from 47 subjects at Boston's Beth Israel Hospital (now the Beth Israel Deaconess Medical Center). Each ECG data sequence has an 11 bit resolution over a 10 mV range with a sampling frequency of 360 Hz. This dataset is fully annotated with both beat-level and rhythm-level diagnoses. 
\item The PhysioNet Computing in Cardiology Challenge 2017 dataset \cite{clifford2017af} (33 papers) contains 8,528 de-identified ECG recordings with durations ranging from 9 s to just over 60 s that were sampled at 300 Hz by an AliveCor healthcare device. Among these recordings, 5154 are normal, 717 recordings are AF, 2,557 recordings are others, and 46 recordings are noise. Additionally, 3,658 test recordings are private for scoring. This dataset was collected by healthcare devices.
\item The PTB Diagnostic ECG Database \cite{bousseljot1995nutzung} (16 papers) contains 549 15 channel ECG records from 290 subjects. The sampling rate reaches as high as 10 kHz. Among these subjects, 216 have one of eight types of heart disease and 52 are healthy control, while 22 are unknown. 
\item The MIT-BIH Atrial Fibrillation Database \cite{moody1983new} (8 papers) includes 25 10 h long-term 2 lead ECG recordings with a sampling rate of 250 Hz for human subjects with AF (mostly paroxysmal). The original recordings were collected at Boston's Beth Israel Hospital using ambulatory ECG recorders with a 0.1 to 40 Hz recording bandwidth. 
\item 2018 The China Physiological Signal Challenge dataset \cite{liu2018open} (seven papers) contains 6,877 (3178 female, 3699 male) 12 lead ECG recordings with durations ranging from 6 s to just over 60 s, which were collected at 11 hospitals with a sampling rate of 500 Hz. Among these recordings, 918 are normal, 1,098 recordings are AF, 704 are first-degree atrioventricular block, 207 recordings are left-bundle branch block (LBBB), 1,695 are right-bundle branch block (RBBB), 556 are PAC, 672 recordings are PVC, 825 are ST segment depression, and 202 are ST segment elevation. Additionally, 2,954 test recordings are private for scoring. 
\end{itemize}

A summary of the deep learning methods tested on the PhysioNet Computing in Cardiology Challenge 2017 dataset is provided in Table \ref{tb:compare_deep}. We only include methods with an $F_{1}$ score over 0.8 reported for the hidden test set. One can find the official leaderboards for the PhysioNet Computing in Cardiology Challenge 2017 at \url{https://physionet.org/content/challenge-2017/1.0.0/}, 2018 China Physiological Signal Challenge at \url{http://2018.icbeb.org/Challenge.html}, and PhysioNet Computing in Cardiology Challenge 2020 at \url{https://physionetchallenges.github.io/2020/}.

\begin{table}
\centering
\resizebox{\columnwidth}{!}{
\begin{tabular}{lc|lllll|l}
\toprule
Method                    & Model         & $F_{1N}$    & $F_{1A}$    & $F_{1O}$    & $F_{1P}$  & $F_{1NAOP}$     & $F_{1NAO}$  \\
\midrule
\cite{teijeiro2018abductive}    & RNN + Expert Features       & 0.9030 & 0.8547 & 0.7366 & 0.5622 & 0.7641 & 0.8314    \\
\cite{hannun2019cardiologist}     & CNN    & 0.91 & 0.84 & 0.74 & NA     & NA    & 0.83  \\
\cite{plesinger2018parallel}     & CNN + Expert Features    & 0.9151 & 0.8247 & 0.7437 & NA     & NA    & 0.8278  \\
\cite{hong2017encase}       & CRNN + Expert Features        & 0.9117 & 0.8128 & 0.7505 & 0.5671 & 0.7605 & 0.8250    \\
\cite{sodmann2018convolutional}  & CNN + Expert Features     & 0.9142 & 0.8153 & 0.7370 & NA     & NA       & 0.8222 \\
\cite{zihlmann2017convolutional} & CRNN   & 0.9090 & 0.8221 & 0.7319 & 0.5676 & 0.7577  & 0.8210           \\
\cite{warrick2018ensembling}     & CRNN         & 0.9028 & 0.8221 & 0.7324 & NA    & NA  & 0.8191    \\
\cite{xiong2018ecg}              & CNN   & 0.9031 & 0.8203 & 0.7310 & 0.5251 & 0.7449 & 0.8181    \\
\cite{parvaneh2018analyzing}     & CNN + Expert Features      & 0.9056 & 0.8284 & 0.7204 & NA     & NA   & 0.8181         \\
\bottomrule
\end{tabular}
}
\caption{Comparison of deep learning methods on the PhysioNet Computing in Cardiology Challenge 2017 dataset. We only include methods with an $F_{1}$ score over 0.8 reported for the hidden test set. }
\label{tb:compare_deep}
\end{table}

\section{Discussion of Opportunities and Challenges}

In this section, we will discuss the current challenges and problems related to deep learning based on ECG data. Additionally, potential opportunities are also identified in the context of these challenges and problems. 

\subsection{Data Collection}

As shown in Table \ref{tb:data}, there is no standard regarding collection procedures. Different studies have used various numbers of leads, durations, sources (subject backgrounds), etc. This makes it difficult to compare results between different datasets fairly. Additionally, high-quality data and annotations are difficult to acquire, so many current works are still using the MIT-BIH Arrhythmia Database, which was collected over 40 y ago. The most recent single-lead PhysioNet Computing in Cardiology Challenge 2017 and 12 lead 2018 China Physiological Signal Challenge used high-quality data, but they both focused on short-term ECG recordings. Researchers would welcome a new high-quality long-term ECG dataset with annotations and such a dataset would certainly inspire new innovative studies. 

\subsection{Interpretability}

Deep learning models are often considered to be black-box models because they typically have many model parameters or complex model architectures, which makes it difficult for a human to understand why a particular result is generated by such a model. This challenge is much more severe in the medical domain because diagnoses without any explanation are not acceptable for medical experts. 

There have been a few works focusing on enhancing the interpretability of ECG deep learning methods. For example, some works have \cite{li2019interpretability,hong2019combining} explicitly added interpretable expert features to deep learning models that can be used for partial interpretation. Others have used multi-level attention weights \cite{hong2019mina} or attribute scores \cite{strodthoff2018detecting} to generate salient maps based on raw ECG data. There have also been several works \cite{shashikumar2018detection,hong2019combining} on generating lower-dimensional embeddings using t-distributed stochastic neighbor embeddings \cite{maaten2008visualizing} to derive interpretable results. 

There are two worthwhile research directions regarding interpretability. The first is how to interpret complex deep learning models using relatively simple models. For example, one can first construct a black-box deep learning model for a specific task, and then construct a separate interpretable simple model that matches the deep learning model's predictions, and then interpret prediction results based on the simple model \cite{Ribeiro0G16,Ribeiro0G18}. The second one is how to construct an interpretable deep model directly. For example, when designing a deep model architecture, one can borrow neuron connection concepts from tree-based models \cite{WangAL17} or add attention mechanisms to hidden layers \cite{ChoiBSKSS16,hong2019mina} because such mechanisms can be more easily understood by humans. 

\subsection{Efficiency}

Because deep models are complex, it is difficult to deploy large models on portable healthcare devices, which is a major obstacle to applying deep learning models in real-world applications. In this context, one promising research direction is the model compression technique. For example, knowledge distillation is commonly used to transform large and powerful models into simpler models with a minor decrease in accuracy~\cite{HintonVD15}. Additionally, one can use quantization, weight sharing, and careful coding of network weights \cite{HanMD15} to compress a large model. 

\subsection{Integration with Traditional Methods}

Most existing deep learning models are trained in an end-to-end manner, making them difficult to integrate with traditional expert-feature-based methods after model training is completed \cite{parvaneh2018electrocardiogram}. 

There are two main research directions for tackling this issue. The first is to use existing expert knowledge to design DNN architectures \cite{HuMLHX16}. For example, the authors of \cite{hong2019mina} proposed guiding multilevel attention weights using expert features for modeling ECG data. The second is to consider deep models as feature extractors and explicitly extract latent embeddings from deep learning models. To some extent, neural networks with expert features represent a step in this direction. Therefore, one can easily combine expert features with deep features and construct traditional machine learning methods on such features. 

\subsection{Imbalanced Labels}

ECG disease labels are very likely to have very biased distributions because most severe diseases occur rarely, but are very important. It is difficult to train an effective deep learning method with a large number of model parameters using small datasets of disease labels. 

There are two main methods for handling this problem. The first is data augmentation, such as data preprocessing using the side-and-cut technique, or generating synthetic training datasets using generative models, such as variational AEs \cite{doersch2016tutorial} or GANs \cite{goodfellow2014generative}. The second is to design new loss functions, such as focal loss \cite{LinGGHD17}, or new model training schema, such as few-shot learning \cite{abs190405046}. 

For example, the method proposed in \cite{zhou2019k} uses a skewness-driven dynamic data augmentation technique to balance data distributions. The authors of \cite{zhou2019beatgan} and \cite{golany2019pgans} deployed GANs to improve classification accuracy and anomaly detection, respectively. The method proposed in \cite{jiang2019novel} eliminates the negative effects of imbalanced data from the perspectives of resampling, data features, and algorithms. 

\subsection{Multimodal Data}

Currently, most works have only considered ECG data for analysis. However, a few works have considered joint analysis with other data sources. For example, the method proposed in \cite{xu2018raim} predicts intensive care unit patient moralities based on a combination of interventions, lab tests, vital signs, and ECG data. The authors of \cite{hammad2018multimodal} designed a multi-modal biometric authentication system using a fusion of ECG and fingerprint data. 

Based on the development of medical and healthcare devices, many vital signs, such as temperature, respiratory rate, and blood pressure can be collected simultaneously with ECG data. However, these data are not always synchronized with ECG timelines and their sampling frequencies vary significantly, meaning they can be regarded as multimodal data. There is a potential opportunity to study how to design models that are capable of utilizing such multimodal data simultaneously to improve task performance compared to models trained on any individual modality. 

\subsection{Emerging Interdisciplinary Studies}

Finally, there have been a few innovative interdisciplinary studies, a few of which are listed below. 
1) Safe driving intensity and cardiac reaction time assessment based on ECG signals \cite{koh2019evaluation}. 
2) Emotion detection based on ECG signals \cite{shu2018review}.
3) Mammalian ECG analysis \cite{behar2018physiozoo}.
4) Estimating age and gender based on ECG data \cite{attia2019age}. 
The key to the success of such studies is adequate data support. 

\section{Conclusion}

In this paper, we systematically reviewed existing deep learning (DNN) methods for ECG data from the perspectives of models, data, and tasks. We found that deep learning methods can generally achieve better performance than traditional methods for ECG modeling. However, there are still some unresolved challenges and problems related to these deep learning methods. Our main contributions are twofold. First, we provided a systematic overview of various deep learning methods that can be employed in real applications. Second, we highlighted some potential future research opportunities.

\section*{Acknowledgement}
This work was partially supported by the National Science Foundation awards IIS-1418511, CCF-1533768 and IIS-1838042, and the National Institute of Health awards NIH R01 1R01NS107291-01 and R56HL138415.

\printcredits
\bibliographystyle{cas-model2-names}
\bibliography{ref}

\end{document}